%% file: paper.tex
\documentclass[]{llncs}

\usepackage{multirow}
\usepackage{subfigure}
\usepackage[utf8]{inputenc}
\usepackage{tabularx}
\usepackage{colortbl}
\usepackage{xcolor}
\usepackage{graphicx}
\usepackage{tabularx}
\usepackage{booktabs}
\usepackage{balance}
\usepackage[plain]{fancyref}
\usepackage{enumitem}
\usepackage{microtype}
\usepackage[font=small,format=plain,labelfont=bf,textfont=it]{caption}
\usepackage{cite}
\usepackage{url}
\usepackage[hidelinks]{hyperref}
\usepackage{breakurl}
\usepackage{cleveref}

\crefformat{section}{\S#2#1#3} 
\crefformat{subsection}{\S#2#1#3}
\crefformat{subsubsection}{\S#2#1#3}
\definecolor{Gray}{gray}{0.9}

\newcommand{\eg}{{\it e.g.}}
\newcommand{\ie}{{\it i.e.}}

\begin{document}

\title{Can Encrypted DNS Be Fast?}

\author{Austin Hounsel\inst{1}\and
Paul Schmitt\inst{1}\and
Kevin Borgolte\inst{2}\and
Nick Feamster\inst{3}}

\authorrunning{A. Hounsel et al.}

\institute{Princeton University, Princeton, NJ 08544, USA \and
TU Delft, 2628 BX Delft, NL \and
University of Chicago, Chicago, IL 60637, USA\\
\email{\{ahounsel,pschmitt\}@cs.princeton.edu}\\
\email{k.borgolte@tudelft.nl}\\
\email{feamster@uchicago.edu}}

\maketitle

%%%%%%%%%%%%%%%%%%%%  START OF DOCUMENT  %%%%%%%%%%%%%%%%%

\begin{sloppypar}
\input{sections/abstract}
\input{sections/introduction}
\input{sections/method}
\input{sections/results}
\input{sections/related}
\label{lastpage}\input{sections/conclusion}

\subsection*{Acknowledgements}

We thank the Federal Communications Commission's Measuring
Broadband America (MBA) program and the associated MBA-Assisted Research
(MARS) Program for assistance in conducting this study, Jason
Livingood and Al Morton for initial study design suggestions, the MBA
collaborative for experiment input, and Sam Crawford from SamKnows with
assistance in measurement implementation and deployment. This research was
funded in part by National Science Foundation Award CNS-1704077 and a Comcast
Innovation Fund.

% \pagebreak
\bibliographystyle{splncs04}
\bibliography{paper,proceedings}
\end{sloppypar}

\end{document}

%% file: sections/abstract.tex
\begin{abstract}
In this paper, we study the performance of encrypted DNS protocols and conventional DNS from thousands of home networks in the United States, over one month in 2020.
We perform these measurements from the homes of 2,693 participating panelists in the Federal Communications Commission’s (FCC) Measuring Broadband America program.
We found that clients do not have to trade DNS performance for privacy.
For certain resolvers, DoT was able to perform \textit{faster} than DNS in median response times, even as latency increased.
We also found significant variation in DoH performance across recursive resolvers.
Based on these results, we recommend that DNS clients (\eg, web browsers) should periodically conduct simple latency and response time measurements to determine which protocol and resolver a client should use.
No single DNS protocol nor resolver performed the best for all clients.
\keywords{DNS \and privacy \and security \and performance.}
\end{abstract}

%% file: sections/introduction.tex
\section{Introduction}\label{sec:intro}
The Domain Name System (DNS) is responsible for translating human-readable domain names (\eg, \texttt{nytimes.com}) to IP addresses.
It is a critical part of the Internet's infrastructure that users must interact with before almost any communication can occur.
For example, web browsers may require tens to hundreds of DNS requests to be issued before a web page can be loaded.
As such, many design decisions for DNS have focused on minimizing the response times for requests.
These decisions have in turn improved the performance of almost every application on the Internet.

In recent years, privacy has become a significant design consideration for the DNS.
Research has shown that conventional DNS traffic can be passively observed by network eavesdroppers to infer which websites a user is visiting~\cite{zhu2015connection,rfc7626}.
This attack can be carried out by anyone that sits between a user and their recursive resolver.
As a result, various protocols have been developed to send DNS queries over encrypted channels.
Two prominent examples are DNS-over-TLS (DoT) and DNS-over-HTTPS (DoH)~\cite{rfc7858,rfc8484}.
DoT establishes a TLS session over port 853 between a client and a recursive resolver.
DoH also establishes a TLS session, but unlike DoT, all requests and responses are encoded in HTTP packets, and port 443 is used.
In both cases, a client sends DNS queries to a recursive resolver over an encrypted transport protocol (TLS), which in turn relies on the Transmission Control Protocol (TCP).
Encrypted DNS protocols prevent eavesdroppers from passively observing DNS traffic sent between users and their recursive resolvers.
% They also enable users to send encrypted DNS queries to third-party recursive resolvers, rather than resolvers operated by their ISPs.
% As such, these protocols prevent ISPs from potentially using DNS traffic to build profiles of the websites that their subscribers are visiting.
From a privacy perspective, DoT and DoH are attractive protocols, providing confidentiality guarantees that DNS lacked.

Past work has shown that typical DoT and DoH query response times are typically marginally slower than DNS~\cite{lu2019end,boettger2019empirical,hounsel2020comparing}.
However, these measurements were performed from university networks, proxy networks, and cloud data centers, rather than directly from homes.
It is crucial to measure DNS performance from home networks in situ, as they may be differently connected than other networks.
An early study on encrypted DNS performance was conducted by Mozilla at-scale with real browser users, but they did not study DoT, and they did not explore the effects of latency to resolvers, throughput, or Internet service provider (ISP) choice on performance~\cite{mozilla-experiment}.
Thus, the lack of controlled measurements prevents the networking community from fully understanding how encrypted DNS protocols perform for real users.

In this work, we provide a large-scale performance study of DNS, DoT, and DoH from thousands of home networks dispersed across the United States. 
We perform measurements from the homes of 2,693 participating panelists in the Federal Communications Commission’s (FCC) Measuring Broadband America program from April~7th, 2020 through May~8th, 2020.
We measure query response times and connection setup times using popular, open recursive resolvers, as well as resolvers provided by local networks.
We also study the effects of latency to resolvers, throughput, and ISP choice on query response times.

%% file: sections/method.tex
\section{Method}\label{sec:measurement_method}
In this section, we describe the measurement platform we used to study DNS, DoT, and DoH performance and outline our analyses.
We then describe the experiments we conduct and their limitations.

\subsection{Measurement Platform}
The FCC contracts with SamKnows~\cite{samknows} to implement the operational and logistical aspects of the Measuring Broadband America (MBA) program~\cite{fcc-mba}.
SamKnows is a company that specializes in developing custom software and hardware (also known as ``Whiteboxes'') to evaluate the performance of broadband access networks.
Whiteboxes act as Ethernet bridges that connect directly to existing modems/routers, which enables us to control for poor Wi-Fi signals and cross-traffic.
In accordance with MBA program objectives, SamKnows has deployed Whiteboxes to thousands of volunteers' homes across the United States.
We were granted access to the MBA platform through the FCC's MBA-Assisted Research Studies program (MARS)~\cite{fcc-mars}, which enables researchers (generally from the United States) to run measurements from the Whiteboxes.
We utilize the platform to evaluate how DNS, DoT, and DoH perform from home networks.

% \subsubsection{DNS Query Tool.}\label{sec:querytool}
We perform measurements from each Whitebox using SamKnows' DNS query tool.
For each query, the tool reports a success/failure status (and failure reason, if applicable), the DNS resolution time excluding connection establishment (if the query was successful), and the resolved record~\cite{samknows-dns}.
For DoT and DoH, the tool separately reports the TCP connection setup time, the TLS session establishment time, and the DoH resolver lookup time.
For this study, we only study queries for 'A' and 'AAAA' records.
We note that queries for DNS and DoT are sent synchronously, \ie, they must each receive a response before the next query can be sent.
On the other hand, DoH queries are sent asynchronously, functionality that is enabled by the underlying HTTP protocol.

The query tool handles failures in several ways.
First, if a response with an error code is returned from a recursive resolver (\eg, NXDOMAIN or SERVFAIL), then the matching query is marked as a failure.
Second, if the tool fails to establish a DoT or DoH connection, then all queries in the current batch (explained in~\Fref{sec:experiment}) are marked as failures.
Third, the query tool times out conventional DNS queries after three seconds, at which point it re-sends them.
If three timeouts occur for a given query, the tool marks the query as a failure.
Finally, the query tool marks DoT/DoH queries as failures if either five seconds have passed or if TCP hits the maximum number of re-transmissions allowed by the operating system's kernel (Linux 4.4.79).
The Whiteboxes we measure use the default TCP settings configured by the kernel (\eg, \textit{net.ipv4.tcp\_frto = 2}, \textit{net.ipv4.tcp\_retries1 = 3}, and \textit{net.ipv4.tcp\_retries2 = 15}).

% \input{figures/Whitebox_locations}

% \Fref{fig:whitebox_locations} shows a map indicating where the Whiteboxes we analyzed are deployed across the United States.
% Each state is colored based on how many devices are deployed there.
In total, we collected measurements from 2,804 Whiteboxes, each of which use the latest generation of hardware and software (8.0)~\cite{samknows-whiteboxes}.
Our measurements were performed continuously from April 7th, 2020 through May 8th, 2020 in collaboration with SamKnows and the FCC.
We filtered out certain Whiteboxes from our analysis in several ways.
First, we filtered out 56 Whiteboxes that we did not have \textit{any} network configuration information about (\eg, ISP speed tier, ISP name, and access technology).
Second, we filtered out 25 Whiteboxes that were connected by satellite.
Third, we filtered out 30 Whiteboxes for which we did not know the access technology or ISP speed tier.
This left us with 2,693 Whiteboxes to analyze, with 96\% of queries marked as successful.
The Whiteboxes were connected to 14 ISPs over cable, DSL, and fiber.
% As shown, the Whiteboxes we analyzed are slightly more concentrated along the East and West coasts compared with the center of the United States.

% \input{tables/successful_measurement_counts}

\subsection{Analyses}
We studied DNS, DoT, and DoH performance across several dimensions: connection setup times, query response times for each resolver and protocol, and query response times relative to latency to resolvers, throughput, and ISPs.
Our analyses are driven by choices that DNS clients are able to make (\eg, which protocol and resolver to use) and how these choices affect DNS performance.

\subsubsection{Connection Setup Times.}
Before any query can be issued for DoT or DoH, the client must establish a TCP connection and a TLS session.
As such, we measure the time to complete a 3-way TCP handshake and a TLS handshake.
Additionally, for DoH, we measure the time to resolve the domain name of the resolver itself.
The costs associated with connection establishment are amortized over many DoT or DoH queries as the connections are kept alive and used repeatedly once they are open.
We study connection setup times in~\Fref{sec:connection}.

\subsubsection{DNS Response Times.}
Query response times are crucial for determining the performance of various applications.
Before any resource can be downloaded from a server, a DNS query often must be performed to learn the server's IP address (assuming a response is not cached).
As such, we study query response times for each resolver and protocol in~\Fref{sec:protocolperformance}.
We remove TCP and TLS connection establishment time from DoT and DoH query response times.
The DNS query tool we use closes and re-establishes connections after ten queries (detailed in~\Fref{sec:measurement_protocol}).
As this behavior is unlikely to mimic that of stub resolvers and web browsers~\cite{stubby,firefox-trr-service-channel,firefox-libpref-init-all}, we remove connection establishment times to avoid negatively biasing the performance of DoT and DoH.

\subsubsection{DNS Response Times Relative to Latency and Throughput.}
Conventional DNS performance depends on latency, as the protocol is relatively lightweight; therefore, latency to the DNS resolver can have a significant effect on overall performance.
Furthermore, encrypted DNS protocols may perform differently than conventional DNS in response to higher latency, as they are connection-oriented protocols. 
We study the effect of latency on query response times for each open resolver and protocol in~\Fref{sec:networkperformance}.
SamKnows also provides us with the subscribed downstream and upstream throughput for each Whitebox. We use this information to study the effect of downstream throughput on query response times in~\Fref{sec:networkperformance}.

\subsubsection{DNS Response Times Relative to ISP Choice.}
Lastly, SamKnows provides us with the ISP for each Whitebox. We study query response times for a selection of ISPs in~\Fref{sec:isp_comparison}.

\subsection{Experiment Design}\label{sec:experiment}
We describe below which recursive resolvers and domain names we perform measurements with and how we arrived at these choices.

\input{tables/latency}

\subsubsection{DNS Resolvers.}
For each Whitebox, we perform measurements using three popular open recursive DNS resolvers (anonymized as X, Y, and Z, respectively\footnote{We anonymize the resolvers as per our agreement with the FCC.}), as well as the recursive resolver automatically configured on each Whitebox (the ``default" resolver).
Typically, the default resolver is set by the ISP that the Whitebox is connected to.
Resolvers X, Y, and Z all offer public name resolution for DNS, DoT, and DoH.
However, the default resolver typically only supports DNS.
As such, for the default resolver, we only perform measurements with conventional DNS.
If a Whitebox has configured Resolver X, Y, or Z as its default resolver, then we leave its default resolver measurements out of our analysis.

In~\Fref{tab:latency}, we include the latency to each resolver across all Whiteboxes.
We measure latency by running five ICMP ping tests for each resolver at the top of each hour and computing the average.
We separate latency to DoH resolvers from latency to DNS and DoT resolvers because the domain names of DoH resolvers must be resolved in advance.
As such, the IP addresses for the DoH resolvers are not always the same as DNS and DoT resolvers.
We note that the latencies for the default resolvers are particularly low because these resolvers are often DNS forwarders configured on home routers.
We exclude measurements with five failures or with an average latency of zero (0.7\% of the total measurements).

We identified 41 Whiteboxes with median latencies to Resolvers X, Y, and Z DNS of up to 100
ms, despite median query response times of less than 1 ms. We
consulted with SamKnows, and based on their experience, they believed this behavior could be attributed to DNS
interception by middleboxes between Whiteboxes and recursive resolvers.
For example, customer-premises equipment (CPE) can run DNS proxies (\eg, dnsmasq)
that can cache DNS responses to achieve such low query response times.
Furthermore, previous reports from the United Kingdom indicate that ISPs can
provide customer-premises equipment that is capable of passively observing and
interfering with DNS queries~\cite{jackson2019firmware}. We found that 29 of
these 41 Whiteboxes are connected to the same ISP. We also
identified two Whiteboxes with median latencies to X, Y, and Z DoH of less than 1 ms. Lastly, we identified one Whitebox with median latencies to X, Y, and Z DoT of up to 100 ms, despite median query response times of less than 1 ms.
We analyze the data for these Whiteboxes for completeness.

\subsubsection{Domain Names.}
Our goal was to collect DNS query response times for domain names found in websites that users are likely to visit.
We first selected the top 100 websites in the Tranco top-list, which averages the rankings of websites in the Alexa top-list over time~\cite{le2019tranco}.
For each website selected, we extracted the domain names of all included resources found on the page.
We obtained this data from HTTP Archive Objects (or ``HARs'') that we collected from a previous study~\cite{hounsel2020comparing}.

Importantly, we needed to ensure that the domain names were not sensitive in nature (\eg, \texttt{pornhub.com}) so as to not trigger DNS-based parental controls.
As such, after we created our initial list of domain names, we used the Webshrinker API to filter out domains associated with adult content, illegal content, gambling, and uncategorized content~\cite{webshrinker}.
We then manually reviewed the resulting list.
In total, our list included 1,711 unique domain names.\footnote{Our list of domain names that we measured is available at \url{https://github.com/noise-lab/dns-mba-public.git}.}

\subsubsection{Measurement Protocol.}\label{sec:measurement_protocol}
% We needed to account for several important considerations when designing our experiment.
% First, we needed to ensure that the volume of queries issued by each Whitebox would not cause firewalls to block subsequent queries.
% We wanted to perform queries for 1,712 unique domain names across a total of ten different resolver/protocol combinations: conventional DNS using the local resolver, plus the combinations of all three DNS protocols and three open recursive resolvers.
% Furthermore, we need to ensure that non-DNS measurements running on the same Whiteboxes had adequate resources.
% 
% Second, we needed to ensure that our schedule for sending DoT and DoH queries was similar to how browsers and stub resolvers send DoT and DoH queries.
% For example, Firefox attempts to maintain a persistent HTTP connection for $\sim$28 minutes in its DoH implementation to amortize connection setup times~\cite{firefox-trr-service-channel,firefox-libpref-init-all}.
% Stubby--a stub resolver based on getdns that supports DoT--attempts to re-use TLS connections for up to 10 seconds by default using the EDNS0 keepalive option~\cite{stubby}.
% As such, our measurements needed to mimic the behavior of popular DNS clients by re-using existing TLS connections.
% 
% Finally, we need to account for any cross-traffic that might be sent on the home gateways that the Whiteboxes are connected to.
% For example, if a high volume of web traffic is traversing the home gateway, then the DNS response times for our measurements may be affected.

The steps we take to measure query response times from each Whitebox are as follows:
\begin{enumerate}
    \item We randomize the input list of 1,711 domain names at the start of each hour.
    \item We compute the latency to each resolver with a set of five ICMP ping tests. 
    \item We begin iterating over the randomized list by selecting a batch containing ten domain names.
    \item We issue queries for all 10 domain names in the batch to each resolver / protocol combination.
          For DoT and DoH, we re-use the TLS connection for each query in the batch, and then close the connection.
          If a batch of queries has not completed within 30 seconds, we pause, check for cross-traffic, and retry if cross-traffic is present.
          If there is no cross traffic, we move to the next resolver/protocol combination.
    \item We select the next batch of 10 domain names. If five minutes have passed, we stop for the hour. Otherwise, we return to step four.
\end{enumerate}

\subsubsection{Limitations.}
Due to bandwidth usage concerns and limited computational capabilities on the Whiteboxes, we do not collect web page load times while varying the underlying DNS protocol and resolver.
% Although the Whitebox platform is technically able to measure page loads, the number of possible combinations of pages, recursive resolvers, and protocols result in experiments that require more than an hour to complete.
% We note that previous work suggests that DNS, DoT, and DoH response times correlate to page load times from cloud data centers~\cite{hounsel2020comparing}.
Additionally, while we conducted our measurements, the COVID-19 pandemic caused many people to work from home.
We did not want to perturb other measurements being run with the Measuring Broadband America platform or introduce excessive strain on the volunteers' home networks.
Due to these factors, we focus on DNS response times.

% The platform itself provides limited coverage of the United States.
% For example, there are not as many volunteers that are running Whiteboxes in the central United States, where performance may be different based on the latency to the recursive resolvers.
% Further, although we were able to collect measurements using 14 ISPs, additional ISPs would allow us to paint a more complete picture of DNS, DoT, and DoH performance.\AH{Since we're no longer doing a per-ISP analysis, does it matter that we only collected measurements using 14 ISPs?}

% \subsection{Ethics.}
% In our study we do not collect any user-generated traffic, and we do not perform human experiments.
% We performed measurements by using home networks as vantage points, rather than studying the traffic patterns of users.
% Further, each participant in the our study was a volunteer.
% Therefore, this study was exempt from IRB.

%% file: tables/latency.tex
\begin{table}[t]
  \footnotesize
  \centering
  \begin{tabularx}{\columnwidth}{Xrrrrr}
    \rowcolor{white}
      \toprule
      &
      & \multicolumn{4}{c}{\textbf{Latency (ms)}}
      \\
      \textbf{Resolver}
      & \textbf{Observations}
      & \textbf{Minimum}
      & \textbf{Median}
      & \textbf{Maximum}
      & \textbf{Std Dev}
      \\
      \midrule
        X DNS and DoT
        & 1,593,506
        & 0.94 
        & 20.38
        & 5,935.80
        & 43.61
        \\
        \rowcolor{Gray}
        X DoH
        & 1,567,337
        & 0.14 
        & 22.75
        & 8,929.88
        & 43.25
        \\
        Y DNS and DoT
        & 1,596,964
        & 2.00 
        & 20.90
        & 9,701.82
        & 46.79
        \\
        \rowcolor{Gray}
        Y DoH
        & 1,552,595
        & 0.14 
        & 20.50
        & 10,516.31
        & 40.68
        \\
        Z DNS and DoT
        & 1,579,605
        & 2.35 
        & 31.41
        & 516,844.73
        & 414.26
        \\
        \rowcolor{Gray}
        Z DoH
        & 1,533,380
        & 0.14 
        & 33.00
        & 9,537.42
        & 41.11
        \\
        Default DNS
        & 2,009,086
        & 0.13 
        & 0.85
        & 8,602.39
        & 22.93
        \\
        \bottomrule
  \end{tabularx}
  \caption{Recursive resolver latency characteristics.}
  \label{tab:latency}
\end{table}

%% file: sections/results.tex
\input{figures/dot_doh_setup_times}

\section{Results}\label{sec:measurement_results}

This section presents the results of our measurements. We organize our results
around the following questions: (1)~How much connection overhead does encrypted DNS
incur, in terms of resolver lookup (in the case of DoH), TCP connect time, and
TLS setup time; (2)~How does encrypted DNS perform versus conventional DNS?;
(3)~How does network performance affect encrypted DNS performance?; and (4)~How does
encrypted DNS resolver performance depend on broadband access ISP?
Our results show that in the case of certain resolvers---to our surprise---DoT
had \textit{lower} median response times than conventional DNS, even as
latency to the resolver increased.  We also found significant variation in DoH
performance across resolvers.  

\subsection{How Much Connection Overhead Does Encrypted DNS Incur?}\label{sec:connection}

We first study the overhead incurred by encrypted DNS protocols, due to their
requirements for TCP connection setup and TLS handshakes.
Before any batch of DoT queries can be issued with the SamKnows query tool, a TCP connection and TLS session must be established with a recursive resolver.
In the case of DoH, the resolver's domain name is also resolved (\eg, \texttt{resolverX.com}).
In \Fref{fig:dot_doh_setup_times}, we show timings for different aspects of connection establishment for DoT and DoH.
The results show that lookup times were similar for all three resolvers (\Fref{fig:doh_resolver_lookup_time}).
This result is expected because the same default, conventional DNS resolver is used to look up the DoH resolvers' domain names;
the largest median DoH resolver lookup time was X with 17.1 ms.
Depending on the DNS time to live (TTL) of the DoH resolver lookup, resolution of the DoH resolver may occur frequently or infrequently.

Next, we study the TCP connection establishment time for DoT and DoH for each
of the three recursive resolvers (\Fref{fig:tcp_connect_time}).
For each of the three individual resolvers, TCP establishment time for DoT and
DoH are similar.
Resolvers X and Y are similar; Z experienced longer TCP connection times.
The largest median TCP connection establishment time across all resolvers and protocols (Resolver Z DoH) was 30.8~ms.

Because DoT and DoH rely on TLS for encryption, a TLS session must be established before use.
\Fref{fig:tls_connect_time} shows the TLS establishment time for the three open resolvers.
Again, Resolver Z experienced higher TLS setup times compared to X and Y.
Furthermore, DoT and DoH performed similarly for each resolver.
The largest median TLS connection establishment time across all recursive resolvers and protocols (Resolver Z DoH) was 105.2~ms.
As with resolver lookup overhead, the cost of establishing a TCP and TLS connection to the recursive resolver for a system would ideally occur infrequently, and should be amortized over many queries by keeping the connection alive and reusing it for multiple DNS queries.

Connection-oriented, secure DNS protocols will incur additional latency, but
these costs can be (and are) typically amortized by caching the DNS name of
the DoH resolver, as well as multiplexing many DNS queries over a single TLS
session to a DoH resolver.  Many browser implementations of DoH implement
these practices. For example, Firefox establishes a DoH connection when the
browser launches, and it leaves the connection
open~\cite{firefox-trr-service-channel,firefox-libpref-init-all}.  Thus, the
overhead for DoH connection establishment in Firefox is amortized over time.

In the remainder of this paper we do not include connection establishment
overhead when studying DNS query response times.  We omit connection
establishment time for the rest of our analysis because the DNS
query tool closes and re-opens connections for each batch of queries.  Thus,
inclusion of TCP and TLS connection overheads may negatively skew query
response times.

\subsection{How Does Encrypted DNS Perform Compared With Conventional DNS?}\label{sec:protocolperformance}

We next compare query response times across each protocol and recursive
resolver.  \Fref{fig:overall_dns_timings} shows box plots for DNS response
times across all Whiteboxes for each resolver and protocol.  ``Default''
refers to the resolver that is configured by default on each Whitebox (which
is typically the DNS resolver operated by the Whitebox's upstream ISP).

\input{figures/overall_dns_timings}

\paragraph{DNS performance varies across resolvers.} First of all,
conventional DNS performance varies across recursive resolvers.  For the
default resolvers configured on Whiteboxes, the median query response time
using conventional DNS is 24.8 ms.  For Resolvers X, Y,
and Z, the median query response times using DNS are 23.2 ms, 34.8
ms, and 38.3 ms, respectively.  Although X performs better than
the default resolvers, Y and Z perform at least 10~ms
slower.  This variability could be attributed to differences in 
deployments between open resolvers.

\paragraph{DoT performance nearly matches conventional DNS.} Interestingly DoT
lookup times are close to those of conventional DNS.  For Resolvers
X, Y, and Z, the median query response times for
DoT are 20.9 ms, 32.2 ms, and 45.3 ms, respectively.  Interestingly, for X and
Y, we find that DoT performs 2.3 ms and 2.6 ms \emph{faster} than
conventional DNS, respectively.  For both of these resolvers, the best median
DNS query performance could be attained using DoT.
Z's median response time was 7~ms slower.
The performance improvement of DoT over conventional DNS in some cases is
interesting because conventional wisdom suggests that the connection overhead
of TCP and TLS would be prohibitive. On the other hand, various factors,
including transport-layer optimizations in TCP, as well as differences in
infrastructure deployments, could explain these discrepancies.
It may also be the case that DoT resolvers have lower query loads than 
conventional DNS resolvers, enabling comparable (or sometimes faster) response times.
Investigating the causes of these discrepancies is an avenue for future work.

\paragraph{DoH response times were higher than those for DNS and DoT.} DoH
experienced higher response times than conventional DNS or DoT, although this
difference in performance varies significantly across DoH resolvers.  For
Resolvers X, Y, and Z, the median query response
times for DoH are 37.7 ms, 46.6 ms, and 60.7 ms, respectively.  Resolver
Z exhibited the biggest increase in response latency between DoH and
DNS (22.4 ms).  Resolver Y showed the smallest difference in
performance between DoH and DNS (11.8 ms).  Median DoH response times between
resolvers can differ greatly, with X DoH performing 23 ms faster
than Z DoH.  The performance cost of DoH may be due to the overhead
of HTTPS, as well as the fact that DoH implementations are still relatively
nascent, and thus may not be optimized.  For example, an experimental DoH
recursive resolver implementation by Facebook engineers terminates DoH
connections to a reverse web proxy before forwarding the query to a
DNS resolver~\cite{doh-nginx}.

\input{figures/median_latency_dns_timings}

\subsection{How Does Network Performance Affect Encrypted DNS
Performance?}\label{sec:networkperformance}

We next study how network latency and throughput characteristics affect the
performance of encrypted DNS.

\input{figures/latency_rtt_regression}
\input{tables/regression_stats}

\paragraph{DoT can meet or beat conventional DNS despite high latencies to
resolvers, offering privacy benefits for no performance cost.}
\Fref{fig:median_latency_dns_timings} shows that DoT can perform better than DNS
as latency increases for Resolvers X and Y; in the case of Resolver
Z, DoT nearly matches the performance of conventional DNS.
We observe similar behavior with the linear ridge regression models shown
in~\Fref{fig:latency_rtt_regression}. 
As discussed in~\Fref{sec:protocolperformance}, these results could be explained 
by transport-layer optimizations in TCP, differences in infrastructure deployments, and 
lower query loads on DoT resolvers compared to conventional DNS resolvers.
% This result can be explained by the fact
% that the cost of symmetric encryption is small compared to network latency.

\paragraph{DoH performs worse than conventional DNS and DoT as
latencies to resolvers increase.} \Fref{fig:median_latency_dns_timings} shows
that DoH performs substantially worse when latency between the client and
recursive resolver is high;
\Fref{fig:latency_rtt_regression} shows a similar result with a ridge regression model.
As discussed in~\Fref{sec:protocolperformance}, this result could be explained
by either HTTPS overhead, nascent DoH implementations and deployments, or
both.

\input{figures/package_down_dns_timings}

\paragraph{Subscribed throughput affects DNS performance.}
\Fref{fig:package_down_dns_timings} shows DNS response times across each of
the open resolvers as well as the default resolver.  We bin the downstream
throughput into four groups using clustering based on kernel density estimation.
The performance for all protocols tends to improve as throughput
increases, with DoH experiencing the most relative improvement. For example,
for users with throughput that is less than 25 Mbps, the median
query response times for Resolver Y DoH and Y DNS are 73.4 ms
and 48.7 ms, respectively.  As throughput increases from 25 Mbs to 400 Mbps,
the median query response times for Y DoH and Y DNS are 41.2
ms and 31.4 ms, respectively.  DoT performs similarly to conventional DNS
regardless of downstream throughput.  Across all groups, the absolute
performance difference between Resolver X DoT and X
DNS by 0.2 ms, 1.9 ms, 0.1 ms, and 1.4 ms, respectively.  For Resolver
Y, DoT again performs faster than DNS in median query response times
when throughput is less than 800 Mbps.  For the three lower throughput groups, Y DoT
performs faster than Y DNS by 1.4 ms, 2.5 ms, and 1.7 ms, respectively.

\subsection{Does Encrypted DNS Resolver Performance Vary Across ISPs?}\label{sec:isp_comparison}

\input{figures/isp_dns_timings}
\Fref{fig:isp_timings} shows how encrypted DNS response times vary across
different resolvers and ISPs. In short, the choice of resolver matters, and
the ``best'' encrypted DNS resolver also may depend on the user's ISP.
For instance, while ISP C is comparable to the other ISPs for
queries sent to Resolver X, ISP C has significantly lower query
response times to Resolver Y, and is one of the poorest performing
ISPs on Resolver Z.  The difference in median query response times
between Resolver X DoH and X DNS was 20.9 ms for
Whiteboxes on ISP D, and 8.9 ms for Whiteboxes on ISP E; for Z DoH,
the difference in median times was 34.5 ms for Whiteboxes on ISP D, and 47.9 ms
for Whiteboxes on ISP E. 

Resolver performance can also differ across ISPs.  For ISP B, the median query
response time for Z DoT is 11.1 ms faster than Z DNS.
However, for ISP C, Z DoT is significantly slower than DNS, with a
difference in median query response times of 30.6 ms.  We attribute this
difference in performance to higher latency to Resolver Z via ISP C.
The median latency to Z DNS and DoT across Whiteboxes on ISP C was 50 ms, compared to 18.5 ms on ISP B.

% \input{figures/peak_offpeak_dns_timings}

% \subsection{DNS Response Times for Peak vs. Off-Peak Hours}\label{sec:peak_offpeak}
% \PS{I'd drop this section for space. Especially if stationarity makes it.}
% Finally, we seek to understand whether network busyness has an effect on DNS performance.
% \Fref{fig:peak_offpeak_dns_timings} compares DNS timings during peak (between 9am and 9pm) and off-peak (between 9pm and 9am) traffic hours for Resolvers X, Y, and Z.
% We note that our measurement campaign was conducted during the COVID-19 pandemic, in which many people were working from home.
% Thus, Internet usage patterns may have been somewhat irregular compared to typical days.
% However, given that our measurements were performed over 32 days, we believe the results normalize over time.

% Across all resolvers, we find that query response times did not significantly change between peak hours and off-peak hours.
% For Resolver Y, the median query response times for DNS, DoT, and DoH changed by at most 1ms during peak and off-peak hours.
% For Z, the median query response times for DNS, DoT, and DoH also changed by at most 1ms.
% Finally, for X, the median query response times for DNS, DoT, and DoH changed by at most 2ms.
% Thus, for the resolvers we measured, network busyness do not appear to be a major factor in determining DNS performance.

%% file: figures/dot_doh_setup_times.tex
\begin{figure*}[t!]
    \centering
    \subfigure[DoH Resolver Lookup]{%
        \includegraphics[width=0.31\textwidth]{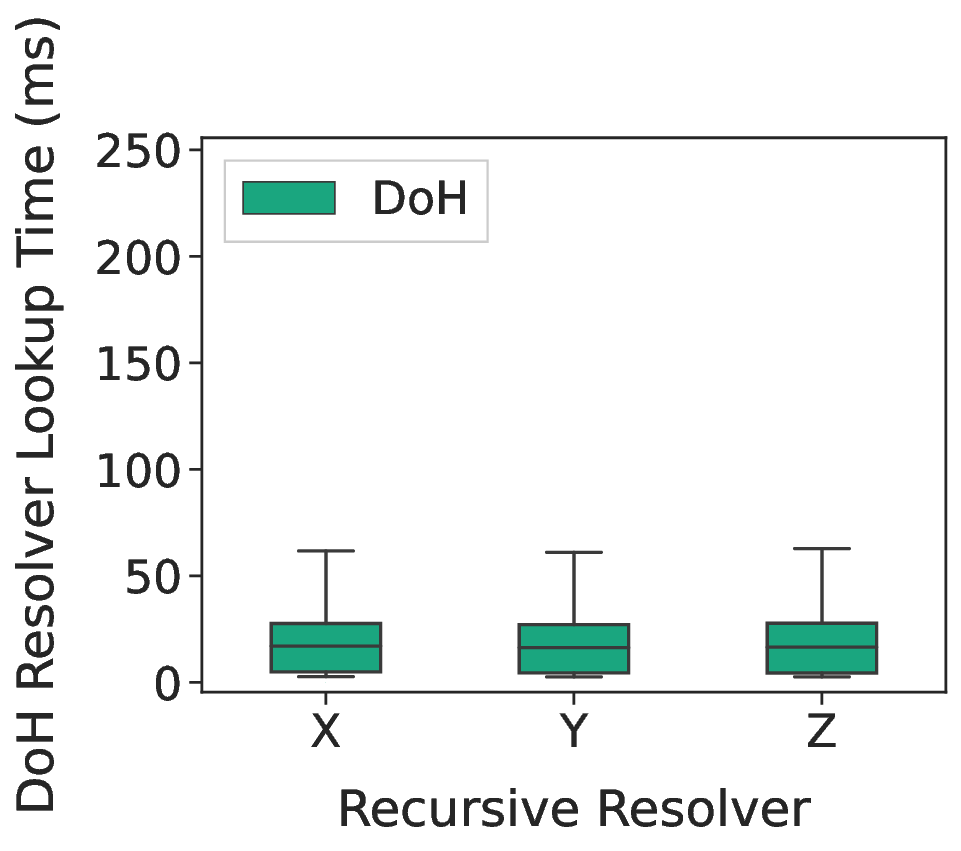}
        \label{fig:doh_resolver_lookup_time}
    }
    \hfill
    \subfigure[TCP Connect Time]{%
        \includegraphics[width=0.31\textwidth]{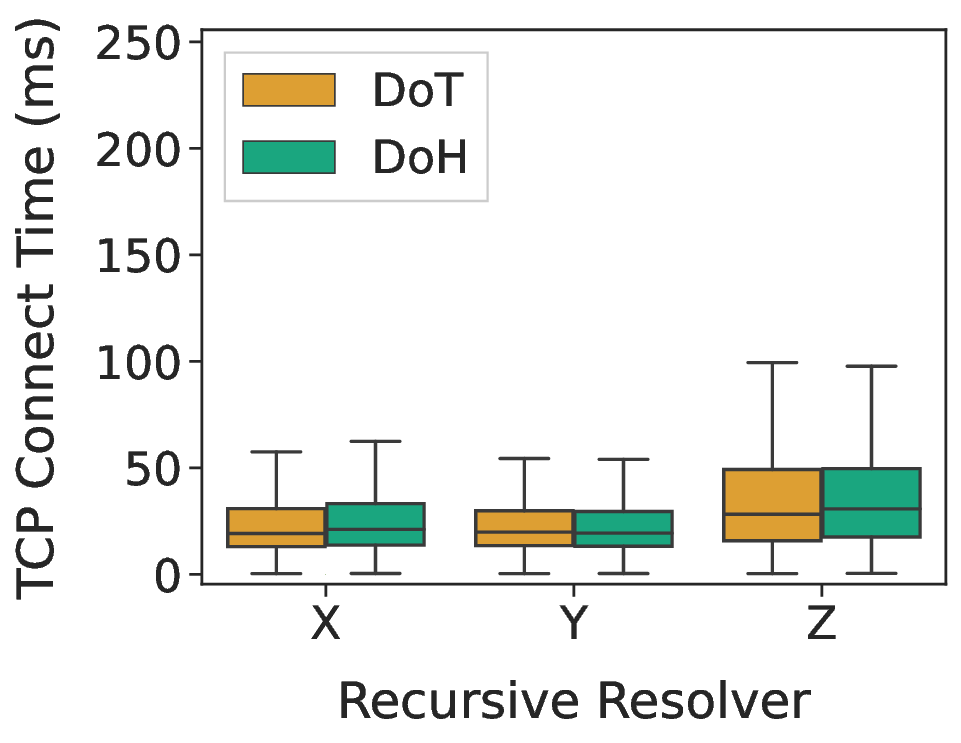}
        \label{fig:tcp_connect_time}
    }
    \hfill
    \subfigure[TLS Setup Time]{%
        \includegraphics[width=0.31\textwidth]{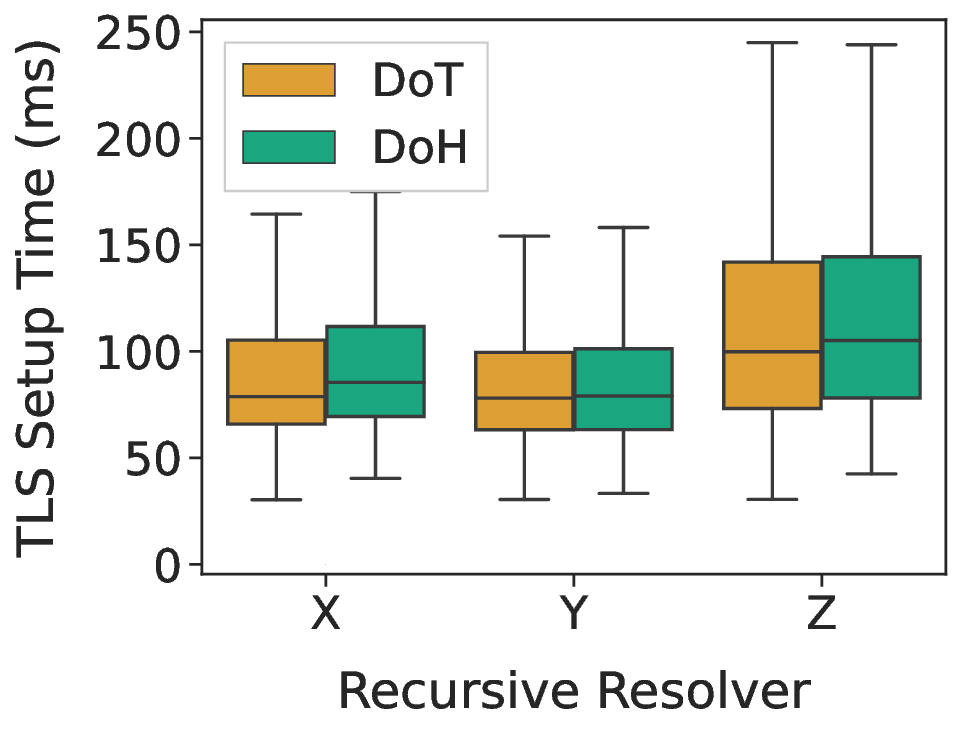}
        \label{fig:tls_connect_time}
    }
    \caption{Connection setup times for DoT and DoH.}
    \label{fig:dot_doh_setup_times}
\end{figure*}

%% file: figures/overall_dns_timings.tex
\begin{figure}[t!]
    \centering
    \includegraphics[width=0.5\linewidth]{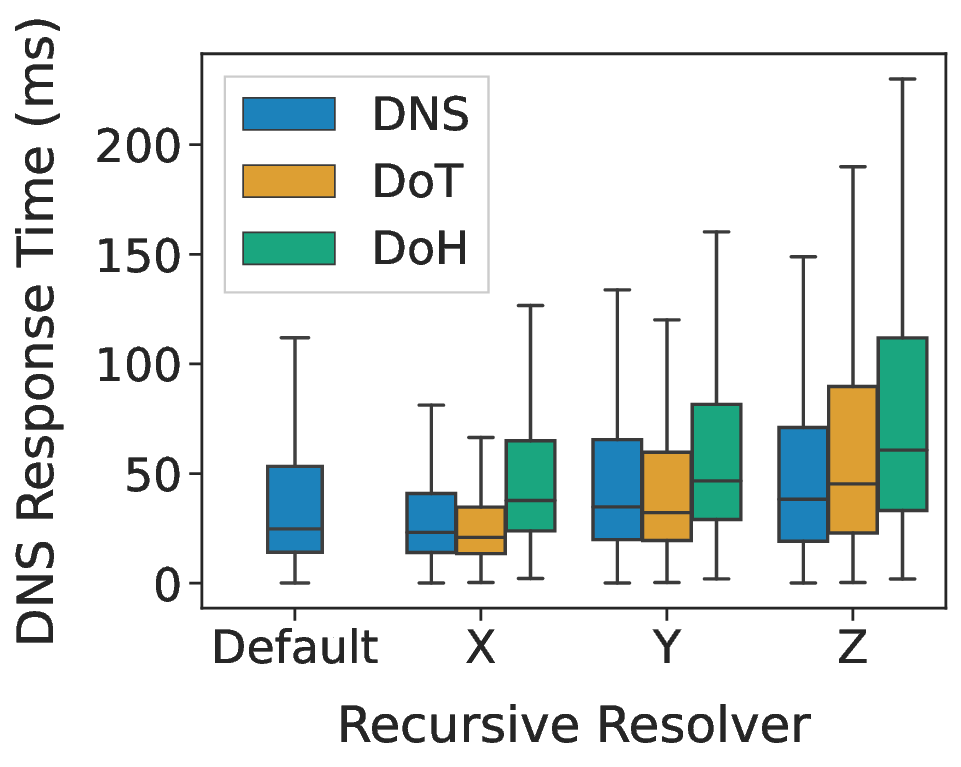}
    \caption{Aggregate query response times.}
    \label{fig:overall_dns_timings}
\end{figure}

%% file: figures/median_latency_dns_timings.tex
\begin{figure*}[t!]
    \centering
    \subfigure[Resolver X]{%
        \includegraphics[width=0.31\textwidth]{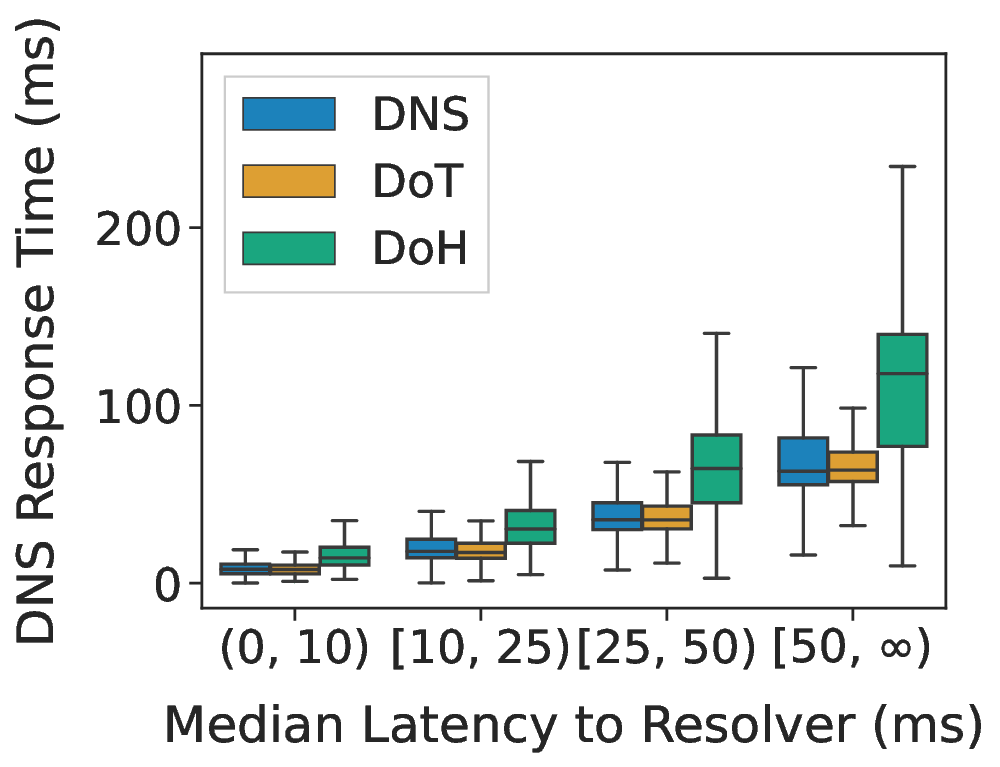}
        \label{fig:median_latency_dns_timings_x}
    }
    \hfill
    \subfigure[Resolver Y]{%
        \includegraphics[width=0.31\textwidth]{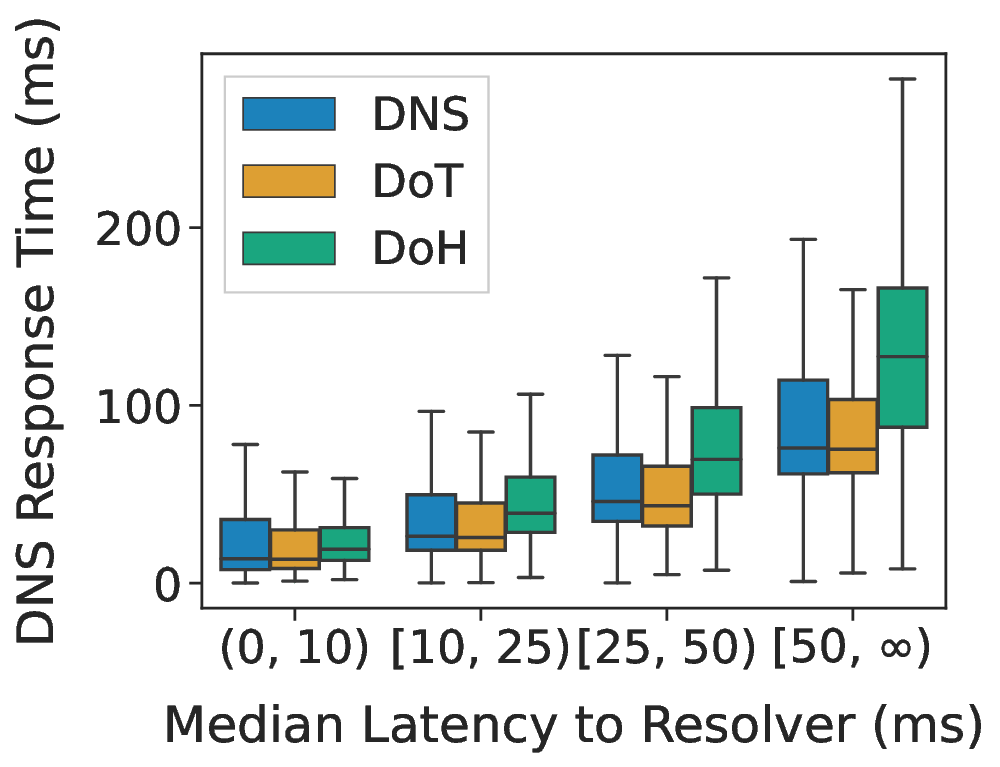}
        \label{fig:median_latency_dns_timings_y}
    }
    \hfill
    \subfigure[Resolver Z]{%
        \includegraphics[width=0.31\textwidth]{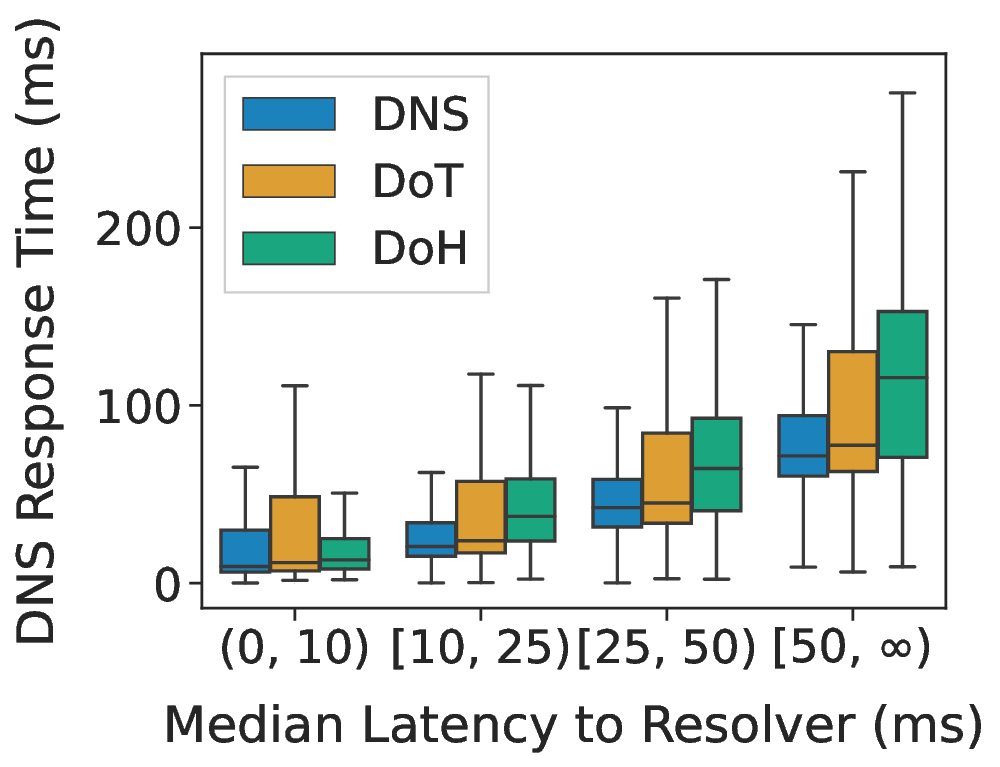}
        \label{fig:median_latency_dns_timings_z}
    }
    \caption{DNS response times based on median latency to resolvers.}
    \label{fig:median_latency_dns_timings}
\end{figure*}

%% file: figures/latency_rtt_regression.tex
\begin{figure*}[t!]
    \centering
    \subfigure[Resolver X]{%
        \includegraphics[width=0.31\textwidth]{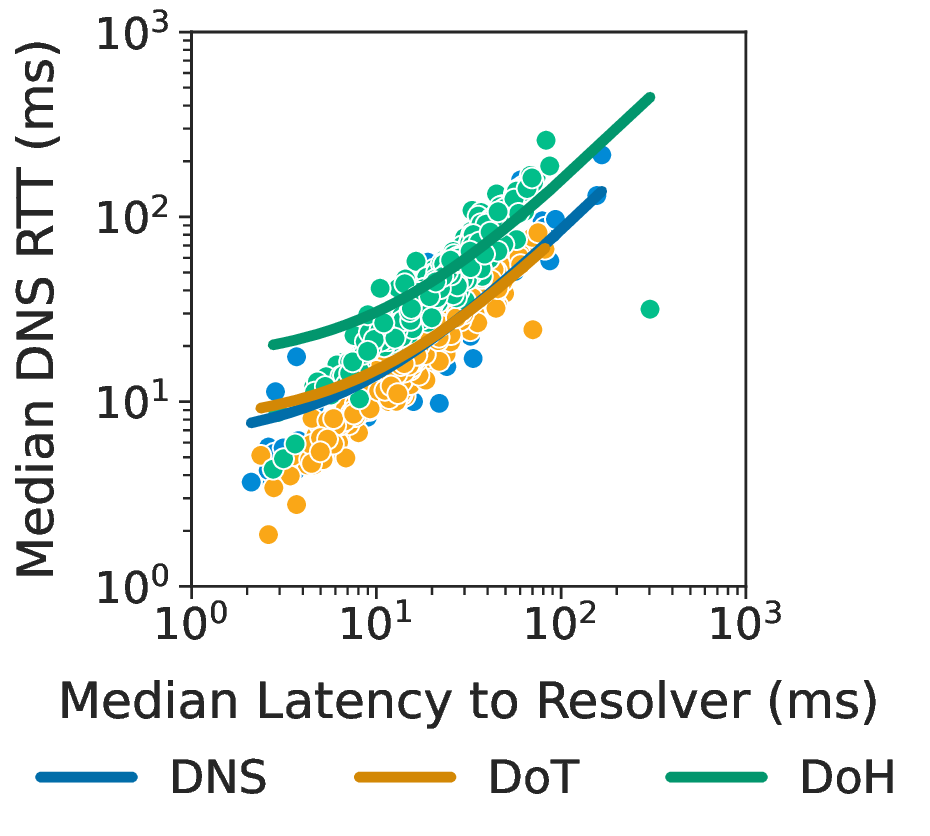}
        \label{fig:latency_rtt_regression_x}
    }
    \hfill
    \subfigure[Resolver Y]{%
        \includegraphics[width=0.31\textwidth]{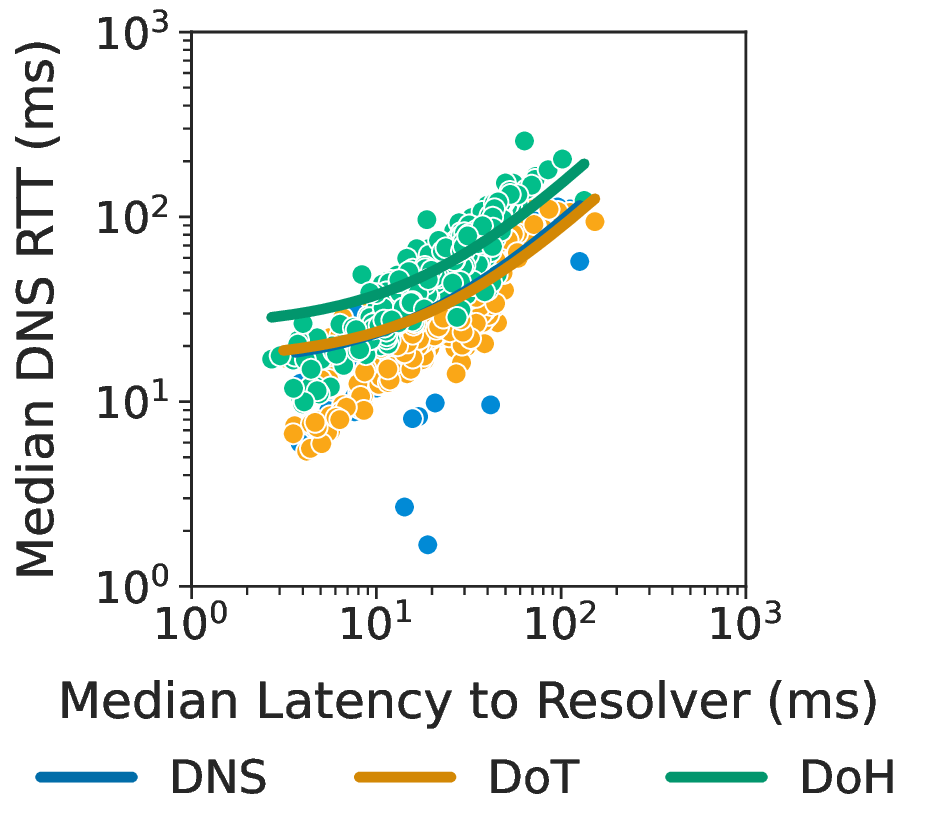}
        \label{fig:latency_rtt_regression_y}
    }
    \hfill
    \centering
    \subfigure[Resolver Z]{%
        \includegraphics[width=0.31\textwidth]{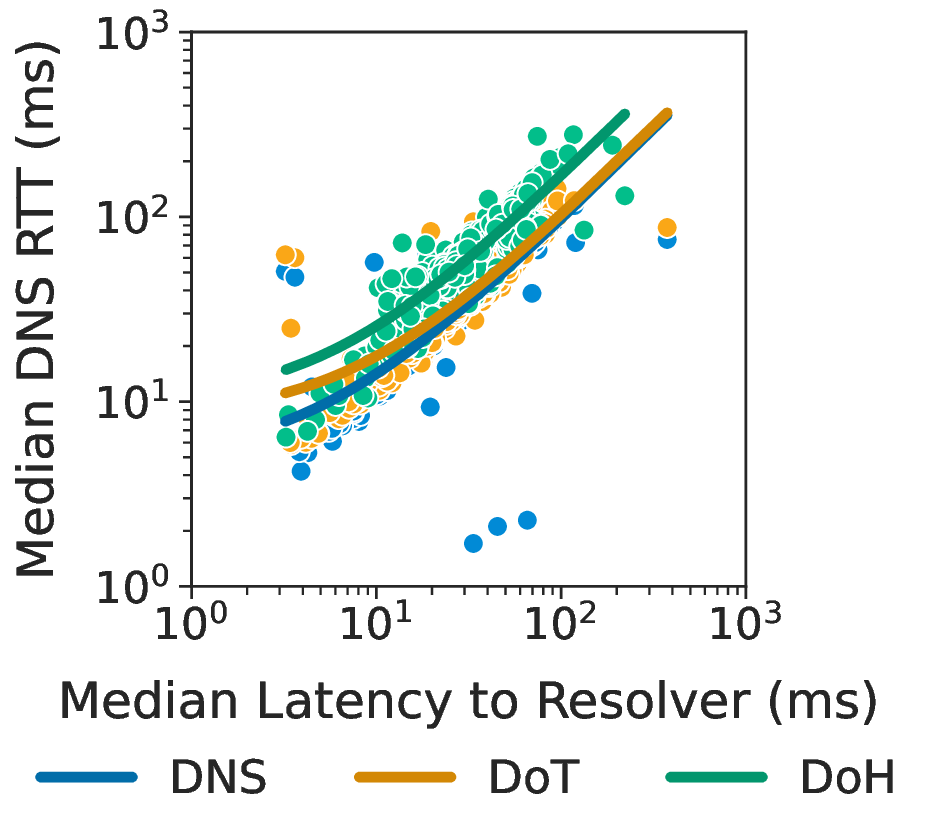}
        \label{fig:latency_rtt_regression_z}
    }
    \caption{Ridge regression models comparing median latency to resolvers to median DNS response times (alpha = 1).}
    \label{fig:latency_rtt_regression}
\end{figure*}

%% file: tables/regression_stats.tex
\begin{table}[t]
  \footnotesize
  \centering
    \begin{tabularx}{\columnwidth}{Xrrrrr}
    \rowcolor{white}
      \toprule
      \textbf{Resolver}
      & \textbf{Coefficient}
      & \textbf{Intercept}
      & \textbf{Mean Absolute Error}
      & \textbf{Mean Squared Error}
      \\
      \midrule
        X DNS
        & 0.79
        & 6.01
        & 3.70
        & 62.06
        \\
        \rowcolor{Gray}
        X DoT
        & 0.74
        & 7.48
        & 4.23
        & 33.89
        \\
        X DoH
        & 1.41
        & 16.39
        & 11.82
        & 551.74
        \\
        \rowcolor{Gray}
        Y DNS
        & 0.79
        & 15.57
        & 8.35
        & 109.25
        \\
        Y DoT
        & 0.71
        & 16.67
        & 9.20
        & 126.43
        \\
        \rowcolor{Gray}
        Y DoH
        & 1.26
        & 25.17
        & 12.36
        & 289.20
        \\
        Z DNS
        & 0.93
        & 4.82
        & 4.46
        & 221.03
        \\
        \rowcolor{Gray}
        Z DoT
        & 0.95
        & 8.07
        & 5.58
        & 221.91
        \\
        Z DoH
        & 1.59
        & 9.75
        & 14.29
        & 482.44
        \\
        \bottomrule
  \end{tabularx}
    \caption{Coefficients, intercepts, and errors for ridge regression models.}
  \label{tab:regression_stats}
\end{table}

%% file: figures/package_down_dns_timings.tex
\begin{figure*}[t!]
    \centering
    \subfigure[Default]{%
        \includegraphics[width=0.45\textwidth]{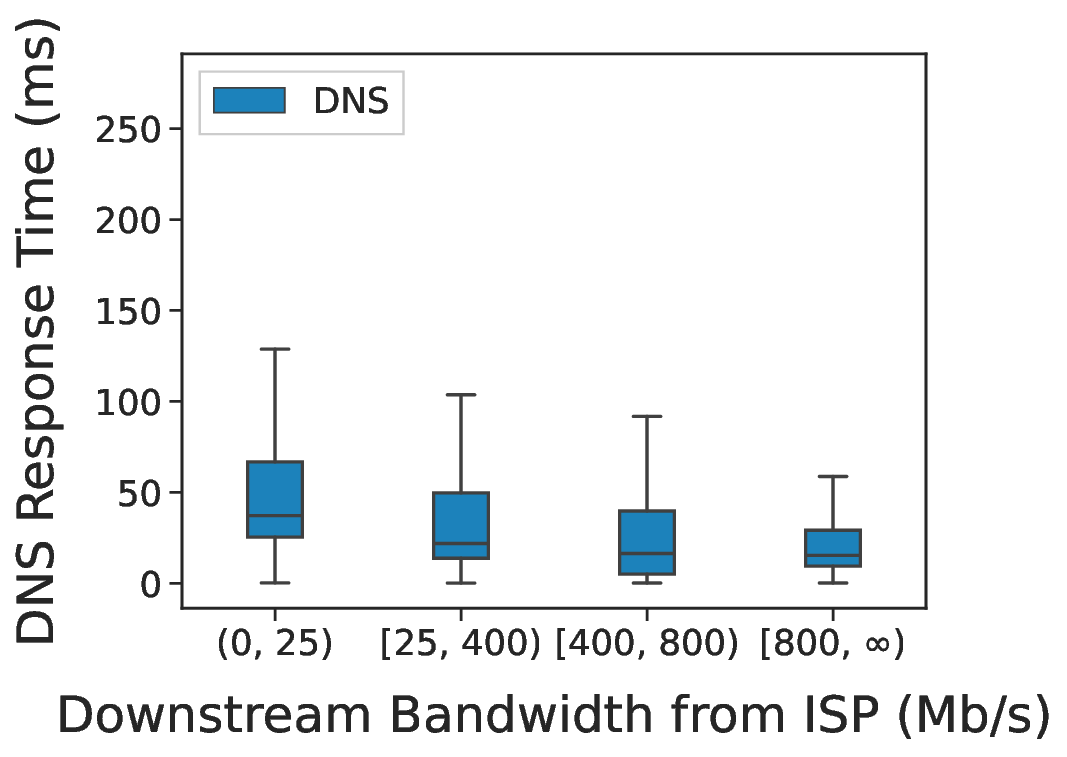}
        \label{fig:package_down_dns_timings_default}
    }
    \subfigure[Resolver X]{%
        \includegraphics[width=0.45\textwidth]{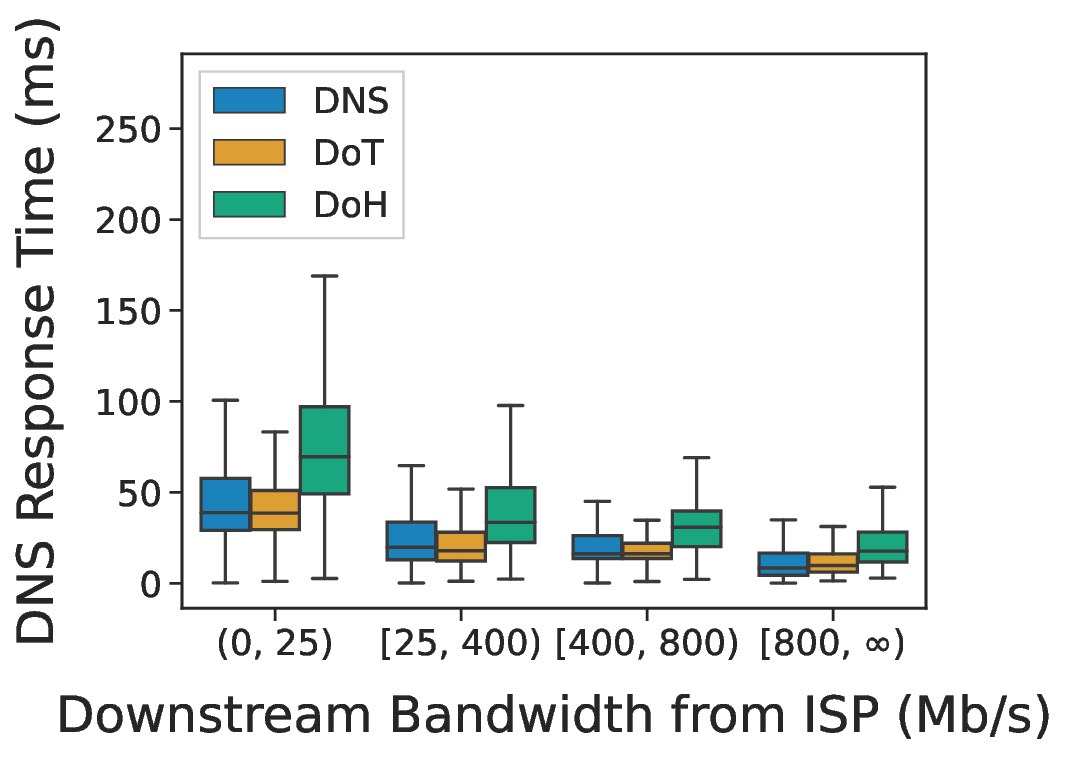}
        \label{fig:package_down_dns_timings_x}
    }
    \\
    \subfigure[Resolver Y]{%
        \includegraphics[width=0.45\textwidth]{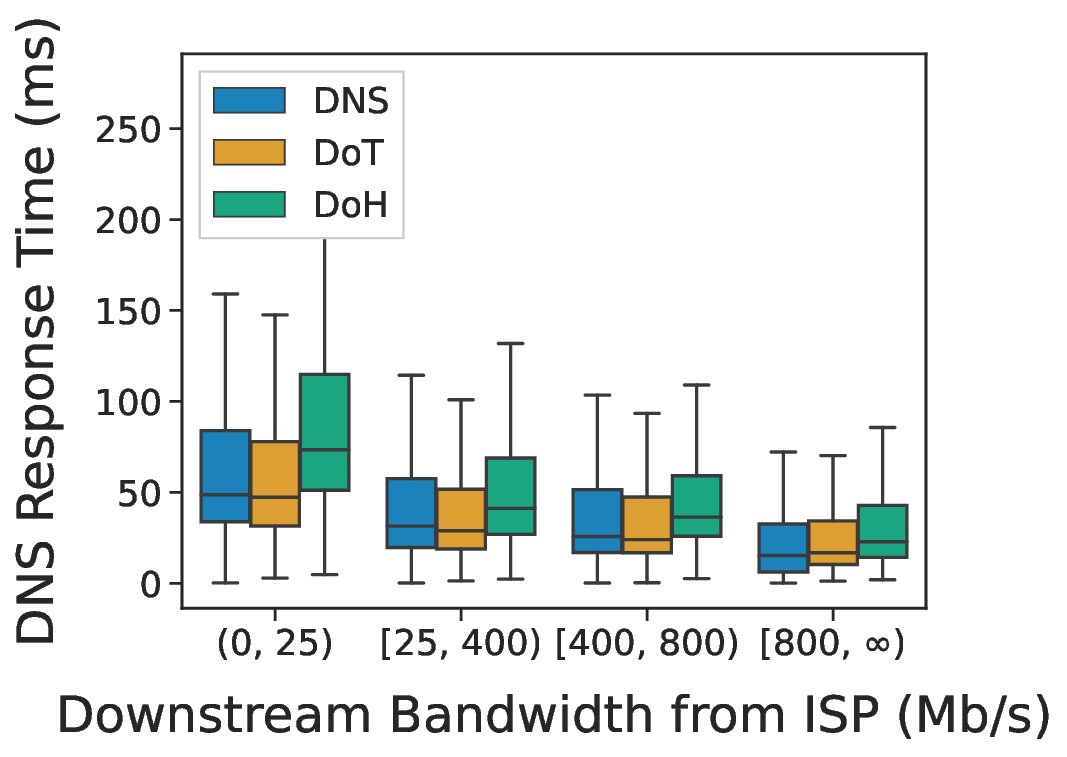}
        \label{fig:package_down_dns_timings_y}
    }
    \subfigure[Resolver Z]{%
        \includegraphics[width=0.45\textwidth]{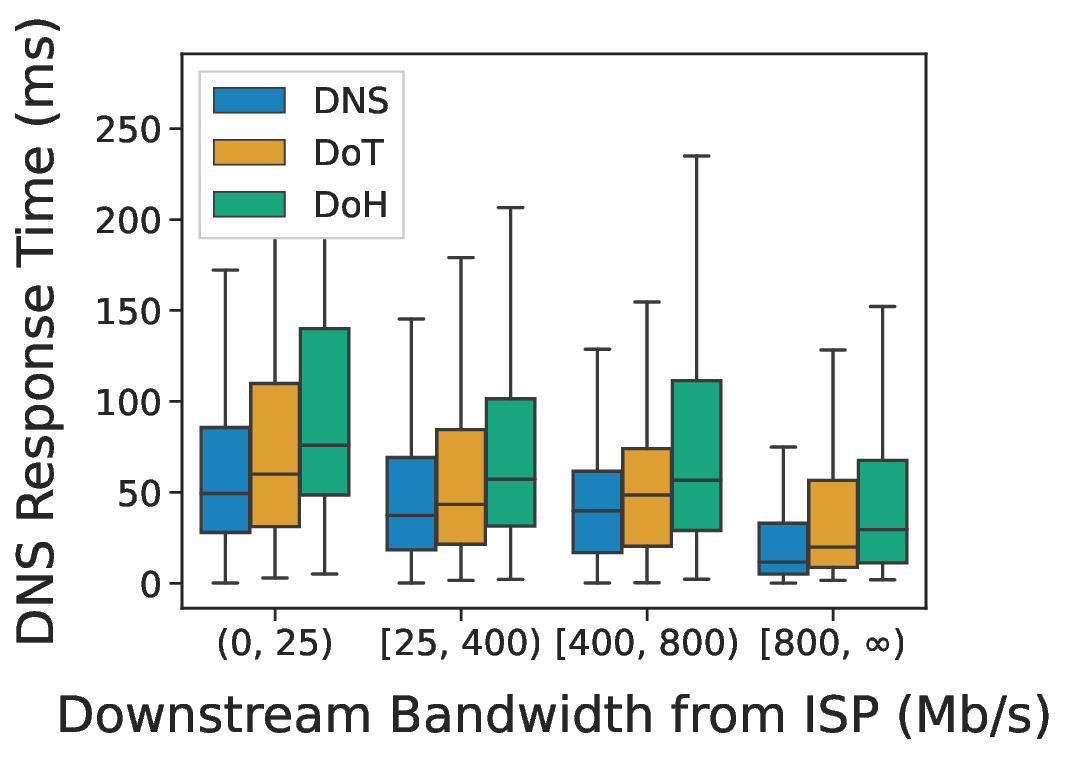}
        \label{fig:package_down_dns_timings_z}
    }
    \caption{Query response times based on downstream access ISP throughput.}
    \label{fig:package_down_dns_timings}
\end{figure*}

%% file: figures/isp_dns_timings.tex
\begin{figure*}[t!]
    \centering
    \subfigure[Resolver X]{%
        \includegraphics[width=0.31\textwidth]{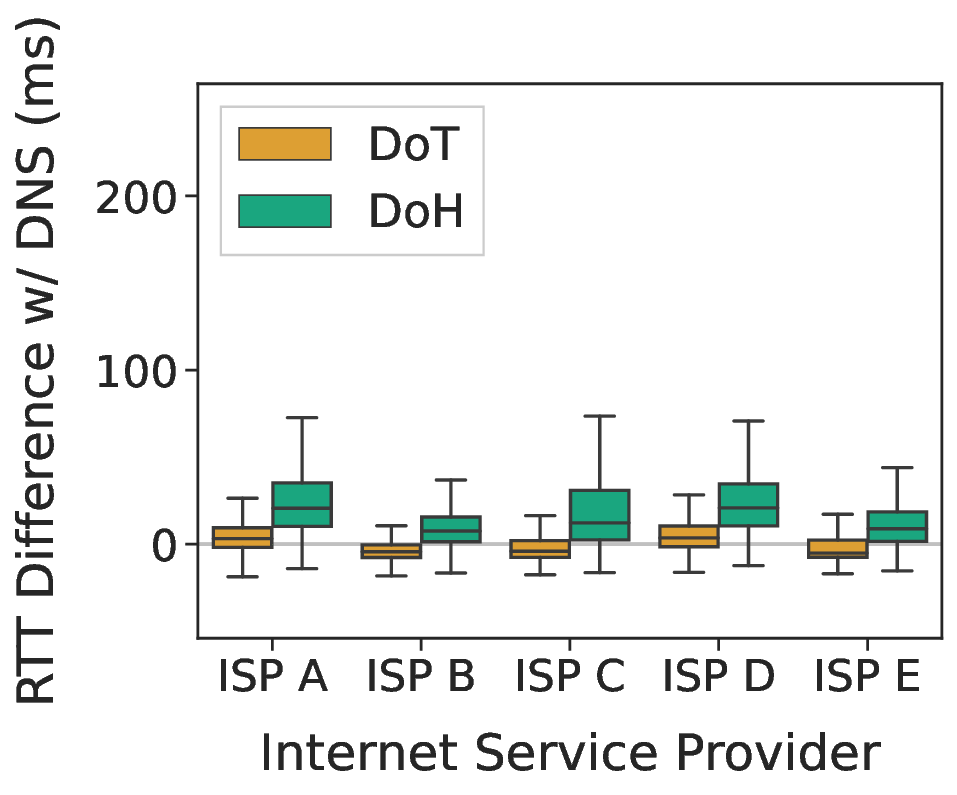}
        \label{fig:isp_timings_x}
    }
    \hfill
    \subfigure[Resolver Y]{%
        \includegraphics[width=0.31\textwidth]{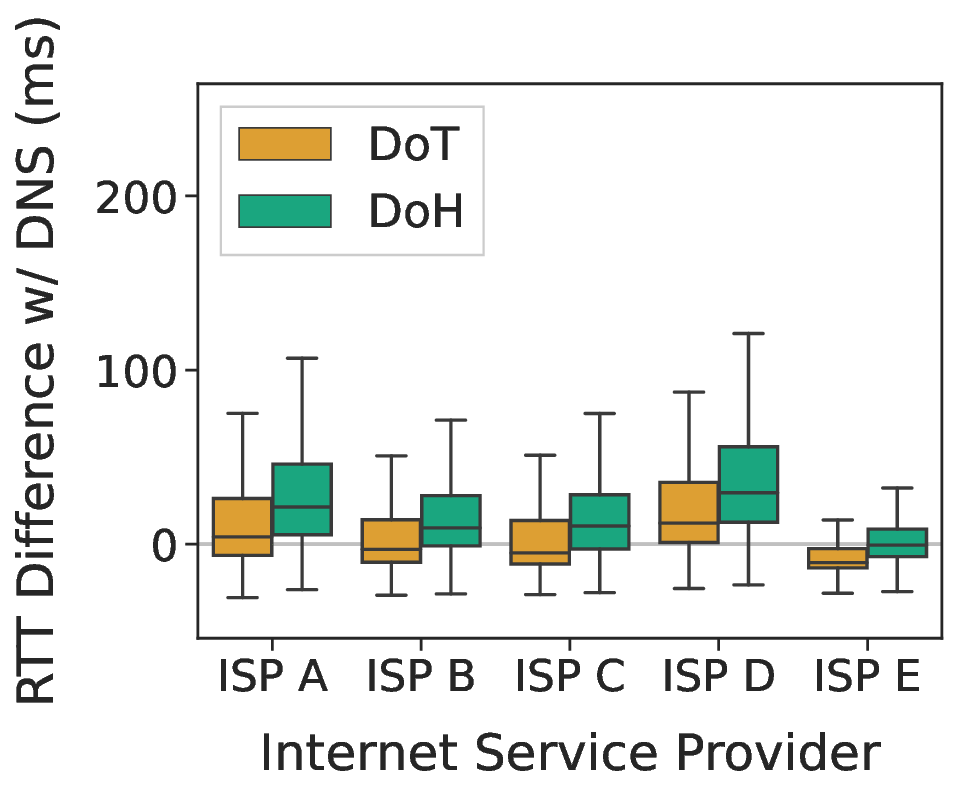}
        \label{fig:isp_timings_y}
    }
    \hfill
    \subfigure[Resolver Z]{%
        \includegraphics[width=0.31\textwidth]{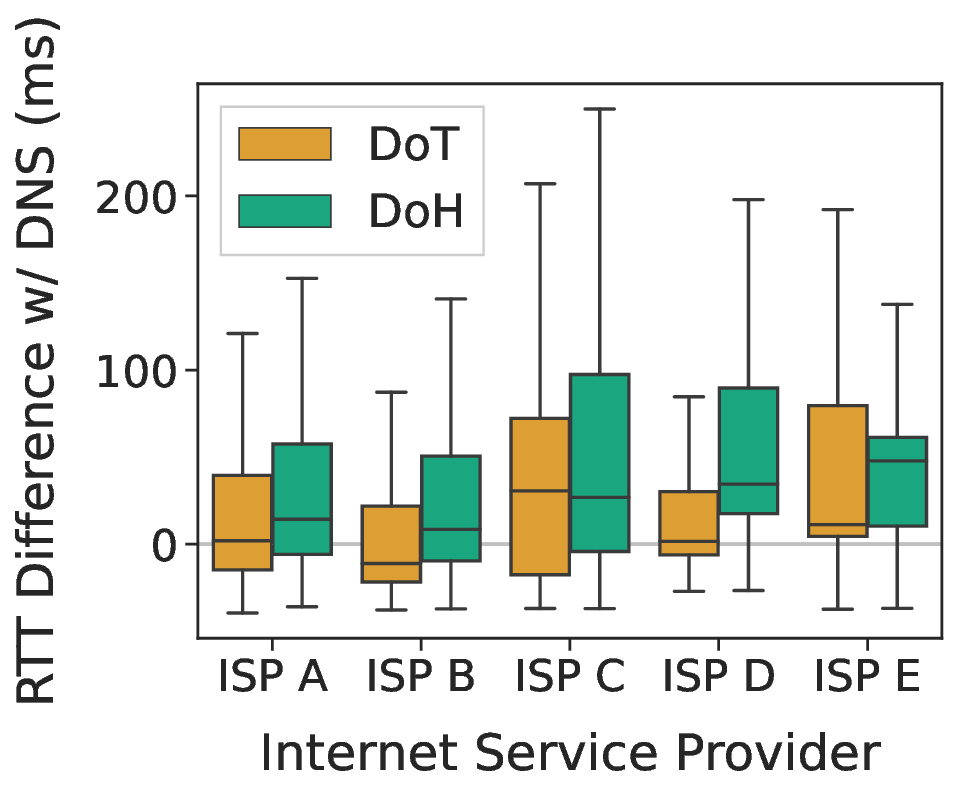}
        \label{fig:isp_timings_z}
    }
    \caption{Per-ISP query response times.}
    \label{fig:isp_timings}
\end{figure*}

%% file: sections/related.tex
\section{Related Work}\label{sec:related}
% In this section, we first compare to related work on the performance of encrypted DNS protocols.
% We then compare to measurements on how DNS impacts web performance.
% Finally, we compare to other studies that conduct measurements from home networks.

% \noindent \textbf{Encrypted DNS Performance.}
Researchers have compared the performance of DNS, DoT, and DoH in various ways.
Zhu et al. proposed DoT to encrypt DNS traffic between clients and recursive resolvers~\cite{zhu2015connection}.
They modeled its performance and found that DoT's overhead can be largely eliminated with connection re-use.
% In our work, we find that DoT be slower or faster than DNS depending on which recursive resolver was used.
% We also performed our measurements from more vantage points than Zhu et al., and we studied the effect of latency and downstream bandwidth on DoT performance.
Böttger et al. measured the effect of DoT and DoH on query response times and page load times from a university network~\cite{boettger2019empirical}.
They find that DNS generally outperforms DoT in response times, and DoT outperforms DoH.
% They also find that much of the performance cost for DoT and DoH can be amortized by re-using TCP connections and TLS sessions.
% However, their methodology relies on collecting HTTP Archive Objects (or "HARs") for query response times, which can contain invalid response times depending on how web page re-directs are triggered.
% This is shown in Figure 6 of their paper, which suggests that for roughly 10\% of websites, the DNS resolution for all included resources can be performed in 0ms.
Hounsel et al. also measure response times and page load times for DNS, DoT, and DoH using Amazon EC2 instances~\cite{hounsel2020comparing}.
% They compare the recursive resolvers for Cloudflare, Google, and Quad9 to the local recursive resolvers provided by Amazon EC2 from five global vantage points in Ohio, California, Seoul, Sydney, and Frankfurt.
% They find that query response times for DoT and DoH are generally slower than DNS.
They find that despite higher response times, page load times for DoT and DoH can be \emph{faster} than DNS on lossy networks.
% However, their measurements were performed from data centers, which may not reflect end-user performance.
Lu et al. utilized residential TCP SOCKS networks to measure response times from 166 countries and found that, in the median case with connection re-use, DoT and DoH were slower than conventional DNS over TCP by 9 ms and 6 ms, respectively~\cite{lu2019end}.

% \noindent \textbf{DNS and Web Performance.}

Researchers have also studied in depth how DNS influences application performance.
Sundaresan et al. used an early MBA deployment of 4,200 home gateways to identify performance bottlenecks for residential broadband networks~\cite{sundaresan2013measuring}.
This study found that page load times for users in home networks are significantly influenced by slow DNS response times.
Wang et al. introduced WProf, a profiling system that analyzes various factors that contribute to page load times~\cite{wang2013demystifying}.
They found that queries for uncached domain names at recursive resolvers can account for up to 13\% of the critical path delay for page loads.
Otto et al. found that CDN performance was significantly affected by clients choosing recursive resolvers that are far away from CDN caches~\cite{otto2012content}.
As a result of these findings. Otto et al. proposed \textit{namehelp}, a DNS proxy that sends queries for CDN-hosted content to directly to authoritative servers.
Allman studied conventional DNS performance from 100 residences in a neighborhood and found that only 3.6\% of connections were blocked on DNS with lookup times greater than either 20 ms or 1\% of the application's total transaction time~\cite{allman2020putting}.

Past work studied the performance impact of ``last mile" connections to home networks in various ways.
Kreibich et al. proposed Netalyzr as a Java applet that users run from devices in their home networks to test debug their Internet connectivity.
Netalyzr probes test servers outside of the home network to measure latency, IPv6 support, DNS manipulation, and more.
Their system was run from over 99,000 public IP addresses, which enabled them to study network connectivity at scale~\cite{kreibich2010netalyzr}.
Dischinger et al. measured bandwidth, latency, and packet loss from 1,894 hosts and 11 major commercial cable and DSL providers in North America and Europe.
This work found that the ``last mile" connection between an ISP and a home network is often a performance bottleneck, which they could not have captured by performing measurements outside of the home network.
However, their measurements were performed from hosts located within homes, rather than the home gateway.
This introduces confounding factors between hosts and the home gateway, such as poor Wi-Fi performance.

%% file: sections/conclusion.tex
\section{Conclusion}\label{sec:conclusion}
In this paper, we studied the performance of encrypted DNS protocols and DNS from 2,693 Whiteboxes in the United States, between April 7th, 2020 and May 8th, 2020.
We found that clients do not have to trade DNS performance for privacy.
For certain resolvers, DoT was able to perform \textit{faster} than DNS in median response times, even as latency increased.
We also found significant variation in DoH performance across recursive resolvers.
Based on these results, we recommend that DNS clients (\eg, web browsers) measure latency to resolvers and DNS response times determine which protocol and resolver a client should use.
No single DNS protocol nor resolver performed the best for all clients.

There were some limitations to our work that point to future research.
First, due to bandwidth restrictions, we were unable to perform page loads from Whiteboxes.
Future work could utilize platforms of similar scale to SamKnows to measure page loads, such as browser telemetry systems.
Second, future work should perform measurements from mobile devices.
DoT was implemented in Android 10, but to our knowledge, its performance has not been studied "in the wild."
Finally, future work could study how encrypted DNS protocols perform from networks that are far away from popular resolvers.
This is particularly important for browser vendors that seek to deploy DoH outside of the United States.